\documentclass[12pt]{article}
\usepackage{svcon2e,graphicx}

\newcommand{\be}{\begin{equation}}
\newcommand{\ee}{\end{equation}}
\newcommand{\ba}{\begin{eqnarray}}
\newcommand{\ea}{\end{eqnarray}}
\newcommand{\ban}{\begin{eqnarray*}}
\newcommand{\ean}{\end{eqnarray*}}

\newcommand{\ket}[1]{\mbox{$ | #1 \rangle $}}

\newcommand{\demi}{\frac{1}{2}}

\newcommand{\one}{\leavevmode\hbox{\small1\normalsize\kern-.33em1}}

\begin{document}

\title{Classical and quantum: some mutual clarifications}
\author{Valerio Scarani}
\date{Group of Applied Physics, University of Geneva\\ 20, rue de
l'Ecole-de-M\'edecine\\ CH-1211 Geneva 4, Switzerland.\\ E-mail:
valerio.scarani@physics.unige.ch} \maketitle

\begin{abstract}
This paper presents two unconventional links between quantum and
classical physics. The {\em first link} appears in the study of
quantum cryptography. In the presence of a spy, the quantum
correlations shared by Alice and Bob are imperfect. One can either
process the quantum information, recover perfect correlations and
finally measure the quantum systems; or, one can perform the
measurements first and then process the classical information.
These two procedures tolerate exactly the same error rate for a
wide class of attacks by the spy. The {\em second link} is drawn
between the quantum notions of "no-cloning theorem" and
"weak-measurements with post-selection", and simple experiments
using classical polarized light and ordinary telecom devices.
\end{abstract}


\section{Introduction}

The boundary between classical and quantum physics is a
fascinating region, that in my opinion, in spite of several
important explorations, has not delivered its deepest treasures. I
will try to motivate this optimistic view on the future of
research in physics by presenting some remarkable links between
"quantum" and "classical" physics.

We have often read in old textbooks or popular books that quantum
physics is the physics of the "infinitely small", while the
"everyday world" is governed by classical physics. This might be
considered, and probably is, a very naive view. Bohr maintained
that the distinction between the classical measurement device and
the quantum measured system is arbitrary but is necessary for our
understanding. The current view of the physicists working in the
field, is that {\em everything is quantum}, the classicality
emerging through interactions (the "everyday world" appears then
to be classical because of the huge amount of interacting
particles involved). This last view, the emergence of classical
behavior simply because of interaction, is nowadays unchallenged
by observation: no phenomenon can be produced as an evidence of
its falseness. Thus, it is a satisfactory description for any
practical purpose, although one may question its validity as a
{\em Weltanschauung}.

The links between "classical" and "quantum" that am I going to
present here are of a different nature: they do not seem to arise
simply from many interacting quantum objects that together exhibit
classical behavior. The first link (Section 2) deals with quantum
cryptography\footnote{This was the topic of my talk during the
Workshop {\em Multiscale Methods in Quantum Mechanics}, Rome,
16-20 December, 2002.}. The second link (Section 3) shows how
typical "quantum" notions (namely, the no-cloning theorem and the
idea of weak measurements with post-selection) manifest themselves
in phenomena that can be described using an entirely classical
theory of light, and that can be revealed using the devices of
ordinary telecommunication networks.

\section{Classical bounds in quantum cryptography}

Quantum cryptography is nowadays the most developed application of
quantum information theory \cite{review}. A more exact name for
quantum cryptography would be {\em quantum key distribution}
(QKD): the goal of the quantum processing is to establish a {\em
secret key} between two distant partners, Alice and Bob, avoiding
the attacks of a possible eavesdropper Eve. Once a common secret
key is established, Alice and Bob will encode the message using
classical secret-key protocols, known to be unbreakable even if
the message is sent on a public authenticated channel.

\begin{figure}[h]
\begin{center}
\includegraphics[width=1\linewidth]{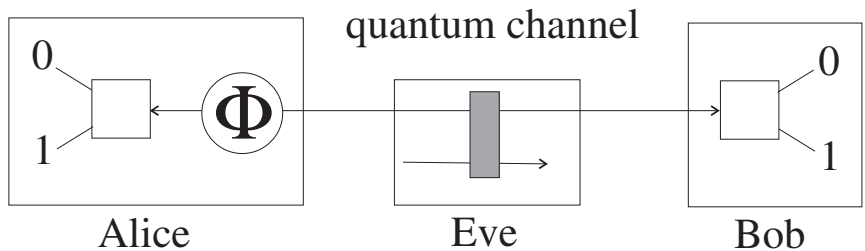}
\caption{The scheme of the QKD implementation with entangled
states. Alice prepares a maximally entangled state $\ket{\Phi}$.
She measures one particle (here, a two-level system) and forwards
the other one to Bob. The spy Eve accesses the quantum channel and
tries to obtain some information by interacting with the flying
particle.} \label{scheme}
\end{center}
\end{figure}

We describe (fig. \ref{scheme}) an implementation of QKD that uses
a source of entangled states, and for clarity we speak of
two-dimensional quantum systems (qubits). Alice has a source that
produce a pair of qubits in the maximally entangled state \ba
\ket{\Phi}_{AB}&=&\frac{1}{\sqrt{2}}\,
\big(\ket{+z}\otimes\ket{+z}+\ket{-z}\otimes\ket{-z}\big)
\nonumber\\&=& \frac{1}{\sqrt{2}}\,
\big(\ket{+x}\otimes\ket{+x}-\ket{-x}\otimes\ket{-x}\big)\,. \ea
She keeps one qubit and forwards the other one to Bob. In the
absence of Eve: (i) if Alice and Bob measure the same observable,
either $\sigma_z$ or $\sigma_x$, they obtain the same result, the
same {\em random bit}; (ii) if one of the partners measures
$\sigma_z$ and the other $\sigma_x$, they obtain completely
uncorrelated random bits. This protocol is repeated a large number
of times. At the end, the items in which Alice and Bob have
performed different measurements are discarded later by public
communication on the classical channel, leaving Alice and Bob with
a list of perfectly correlated random bit: the secret key. This is
what happens in the absence of the spy.

Eve can in principle do whatever she wants on the quantum channel.
The security of QKD comes from the fact that, since any
measurement or interaction perturbs the state, Eve's intervention
cannot pass unnoticed: Alice and Bob know that someone is spying.
Two situations are then possible. (I) Eve has got a "small" amount
of information; in this case, Alice and Bob can process their data
in order to obtain a shorter but completely secret key. Such
classical protocols are the object of important studies in
classical information theory. (II) Eve has got "too much"
information; then Alice and Bob discard the whole key. This may
seem a failure, but it is not: it simply means that the spy has no
other alternative than cutting the channel and forbid any
communication; and this {\em achieves} the goal of cryptography,
because no encrypted message is ever sent that the spy could
decode.

It is then important to quantify the words "small" and "too much"
in the previous discussion: what is the amount of Eve's
information that Alice and Bob can tolerate, that is, at what
critical value are they obliged to discard the whole key? Here is
where remarkable links appear between classical and quantum
information.

\begin{figure}[h]
\begin{center}
\includegraphics[width=0.8\linewidth]{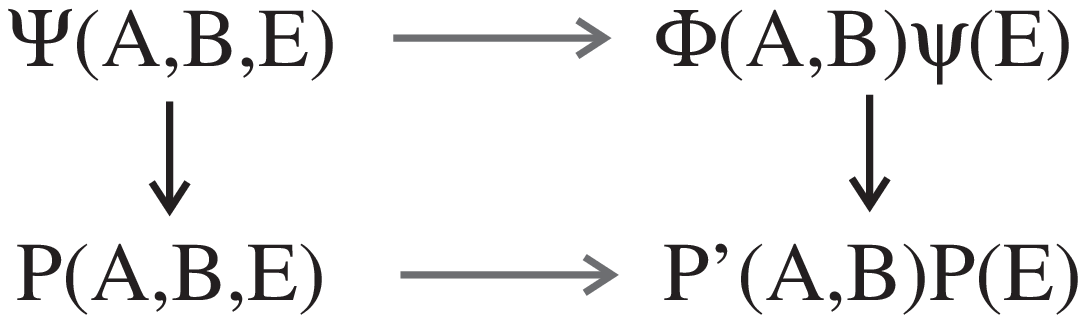}
\caption{Possible ways for the extraction of a secret key. One
starts from a global quantum state $\Psi(A,B,E)$ of Alice, Bob and
Eve, and wants to end up with a classical secret key $P'(A,B)P(E)$
with $P'(A=B)=1$. Grey arrows (horizontal): distillation, quantum
or classical; black arrows (vertical): measurement of the quantum
system, leading to a classical probability distribution.}
\label{diagram}
\end{center}
\end{figure}

We refer to fig. \ref{diagram}. Because of Eve's intervention,
before any measurement the quantum system is in a three-party
entangled state of Alice-Bob-Eve, $\Psi_{ABE}$. Alice and Bob on
their own share the mixed state $\rho_{AB}$, obtained from
$\Psi_{ABE}$ by partial trace on Eve's system. Two procedures are
then possible:

(a) The one that we described above: all the partners make a
measurement, ending in a classical probability distribution
$P(A,B,E)$. Then, Alice and Bob apply classical protocols ({\em
advantage distillation}) in order to extract a shorter secret key,
that is, a shorter list of bits distributed according to a new
distribution $P'(A,B)P'(E)$ in which Eve is uncorrelated and
$P'(A=B)=1$.

(b) If the state $\rho_{AB}$ is entangled, Alice and Bob can delay
any measurement and process many copies of $\rho_{AB}$, to obtain
a smaller number of copies of $\ket{\Phi}_{AB}$ --- and in this
case, automatically Eve is uncorrelated. This procedure is known
as {\em entanglement distillation}, and is one of the fundamental
processes of quantum information. Once Alice and Bob have
$\ket{\Phi}_{AB}$, the measurement provide them immediately with
the secret key.

Having understood this, we can state the main results that have
been obtained:

\begin{itemize}
\item Classical advantage distillation of $P(A,B,E)$ is possible
for bits if and only if quantum entanglement distillation is
possible for the state $\rho_{AB}$ (which is equivalent of asking
that $\rho_{AB}$ is entangled in the case of qubits). This was
demonstrated by Gisin and Wolf when Eve uses the so-called optimal
individual attack \cite{wolf}, and has been recently extended to
all individual attacks \cite{toni}.

\item The same holds for dits ($d$-valued random variables) and
qudits ($d$-dimensional quantum systems), under Eve's individual
attack that is supposed to be the optimal one \cite{us,bruss}. The
demonstration is more involved because not all entangled states of
two qudits are distillable.

\item If $\rho_{AB}$ is entangled enough to violate a Bell
inequality, then a secret key can be extracted from $P(A,B,E)$ in
an {\em efficient} way, that is, using only one-way communication.
This was first proven in Ref. \cite{fuchs}; for the
state-of-the-question, see \cite{us}. A similar result holds for a
multi-partite scheme of key distribution known as "quantum secret
sharing" \cite{scagi}.

\end{itemize}

Mainly because Eve's optimal attack is not generally known, there
are still several open questions. The most important ones are
reviewed in the last section of Ref. \cite{us}.

This concludes my first "unconventional" link between the
classical and the quantum worlds: at the level of information
processing, specifically of the extraction of a secret key from an
initially noisy distribution/state, the critical parameters are
exactly the same, irrespective whether the purification of the
correlations is performed at the quantum or at the classical
level. Moreover, a typically quantum feature such as the violation
of Bell's inequalities is related to the efficiency of the
classical key-extraction procedure.

\section{Quantum physicists meet telecom engineers}

This section is devoted to another kind of unconventional link
between the classical and the quantum world. I prefer let the
examples speak first and draw my conclusions later.

\subsection{No-cloning theorem}

The first example concerns the no-cloning theorem, a well-known
primitive concept of quantum information \cite{noclon}. In its
basic form, it states that no evolution (or more generally, no
trace-preserving completely positive map) can bring
$\ket{\psi}\otimes\ket{0}$ onto $\ket{\psi}\otimes\ket{\psi}$ for
an unknown state $\ket{\psi}$.

\begin{center}
\begin{figure}
\includegraphics[width=1\linewidth]{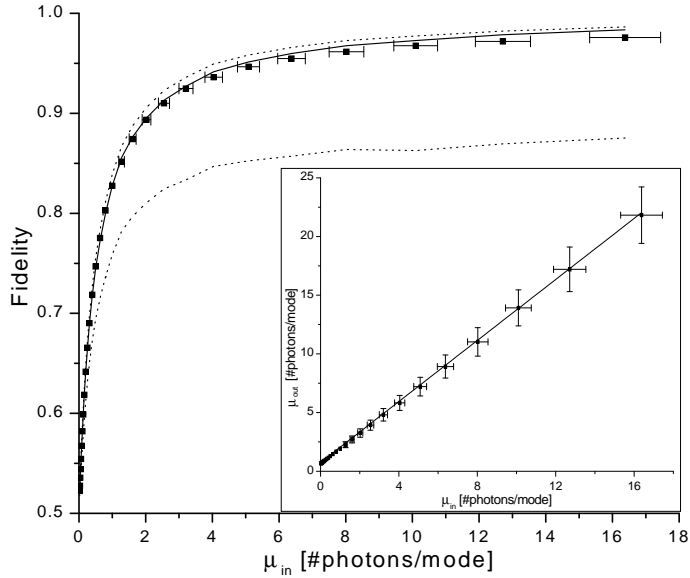}
\caption{Inset: $\mu_{out}$ as a function of $\mu_{in}$; the
linear fit shows that we are far from the saturation of the
amplifier. Main figure: fidelity as a function of $\mu_{in}$.
Solid line: $Q=0.8$, best fit with eq. (\ref{formula}). Dotted
lines: upper: $Q=1$ (optimal cloning); lower: $Q=0$ (no cloning).
From Ref. \cite{fasel}.} \label{results}
\end{figure}
\end{center}

This no-go theorem has motivated the search for an {\em optimal
quantum cloner}: given that perfect cloning is impossible, what is
the best one can do? Optimal cloners have indeed been found and
widely studied; all the meaningful references can be found in any
basic text on quantum information, e.g. \cite{livre}. In the
course of these investigations, a sharp link has been found
between optimal cloning and the well-known phenomenon of
amplification of light: stimulated emission of light in a given
mode (perfect amplification, or cloning) cannot be done without
spontaneous emission (random amplification). Suppose that $N$
photons enter an amplifier, and at the output one selects the
cases in which exactly $M>N$ photons are found: it turns out that
this process {\em realizes the optimal quantum cloning} from $N$
to $M$ copies. The {\em fidelity} of the amplification is the
ratio between the mean number of photons found in the initial mode
(i.e. the mean number of correct copies) and the total number of
copies, $M$ here. The optimal fidelity is found to be \ba
{\cal{F}}_{N\rightarrow M}^{opt}&=&\frac{MN+M+N}{M(N+2)}\,.
\label{fidopt}\ea We realized an experimental demonstration of
optimal cloning using the principle just described to clone the
polarization of light \cite{fasel}. Polarized light of intensity
$\mu_{in}$ is sent into a conventional fiber amplifier (exactly as
those that are used in telecommunications); at the output, we have
an intensity $\mu_{out}$; we separate the input polarization mode
from its orthogonal, and measure the fidelity. The theoretical
prediction for this experiment is \ba \bar{{\cal{F}}}_{\mu_{in}
\rightarrow \mu_{out}} &=&\frac{Q\mu_{out}\,\mu_{in}
+\mu_{out}+\mu_{in}}{Q\mu_{out}\mu_{in}+2\mu_{out}}
\label{formula}\ea where $Q\in [0,1]$ is a parameter related to
the quality of the amplification process. The experimental results
are in excellent agreement (fig. \ref{results}).

It is striking to notice that for $Q=1$, formula (\ref{formula})
is exactly the same as (\ref{fidopt}). Its meaning is however
rather different. In our experiment, {\em everything is
classical}: the laser light is in a coherent state, therefore it
can be described by a classical field; the amplifier is
"classical" in the sense that it transforms coherent states into
coherent states. The quantities $\mu_{in}$ and $\mu_{out}$ that
appear in eq.(\ref{formula}) are not photon numbers as the $N$ and
$M$ of eq. (\ref{fidopt}), but mean values, that have been
measured using an intensity detector. As I said above, all the
devices (source, fibers, amplifier, detectors) are typical of
telecom engineering.

\subsection{Weak measurements with post-selection}

The second example is related to the meaning and physics of the
{\em measurement process}, a widely debated topic of the
foundations of quantum mechanics. In this context, Aharonov,
Vaidman and others introduced the notion of {\em weak measurement
with post-selection} \cite{aav}, sometimes called the "two-state
formalism" of quantum mechanics. The authors' intention in
studying this formalism is strongly motivated by interpretational
issues; that is why most physicists tend to look at these concepts
as artificial ones, introduced on purpose, and that do not add
anything to physics itself. To date, apart from some experiments
that were designed on purpose, only some complex tunnelling
phenomena had received some clarification through this formalism.

We have found however \cite{brunner} that this formalism does
apply to something that exists and is extremely widespread: once
again, the optical telecommunication network. Telecom engineers
are performing weak measurements with post-selection in basically
all that they do! A modern optical network is composed of
different devices connected through optical fibers. With respect
to polarization, two main physical effects are present. The first
one is {\em polarization-mode dispersion} (PMD): due to
birefringency, different polarization modes propagate with
different velocities; in particular, the fastest and the slowest
polarization modes are orthogonal. PMD is the most important
polarization effect in the fibers. The second effect is {\em
polarization-dependent loss} (PDL), that is, different
polarization modes are differently attenuated. PDL is negligible
in fibers, but is important in devices like amplifiers,
wavelength-division multiplexing couplers, isolators, circulators
etc.

\begin{center}
\begin{figure}
\includegraphics[width=0.8\linewidth]{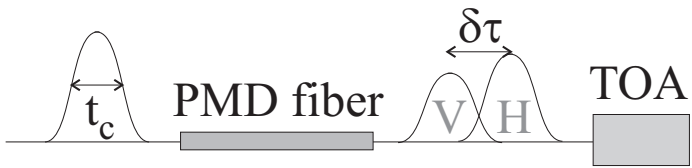}
\caption{When a polarized pulse passing through a PMD fiber, the
polarization mode $H$ (parallel to the birefringency axis in the
Poincar\'e sphere) and its orthogonal $V$ are separated in time. A
measurement of the time-of-arrival (TOA) is a measurement, strong
or weak, of the polarization.}\label{figpmd}
\end{figure}
\end{center}

The first piece of the connection we want to point out is the
following: {\em a PMD element performs a measurement of
polarization on light pulses} (Fig. \ref{figpmd}). In fact, PMD
leads to a separation $\delta\tau$ of two orthogonal polarization
modes in time. If $\delta\tau$ is larger than the pulse width
$t_c$, the measurement of the time of arrival is equivalent to the
measurement of polarization --- PMD acts then as a "temporal
polarizing beam-splitter". However, in the usual telecom regime
$\delta\tau$ is much {\em smaller} than $t_c$. In this case, the
time of arrival does not achieve a complete discrimination between
two orthogonal polarization modes anymore; but still, some
information about the polarization of the input pulse is encoded
in the modified temporal shape of the output pulse. We are in a
regime of {\em weak measurement} of the polarization. The formulae
introduced by Aharonov and co-workers are recovered by measuring
the {\em mean time of arrival}, that is, the "center of mass" of
the output pulse.

The second piece of the connection defines the role of PDL: {\em a
PDL element performs a post-selection of some polarization modes}.
Far from being an artificial ingredient, post-selection of some
modes is the most natural situation in the presence of losses: one
does always post-select those photons that have not been lost!
This would be trivial physics if the losses were independent of
any degree of freedom, just like random scattering; but in the
case of PDL, the amount of losses depends on the meaningful degree
of freedom, polarization. An infinite PDL would correspond to the
post-selection of a precise polarization mode (a pure state, in
the quantum language); a finite PDL corresponds to post-selecting
different modes with different probabilities (a mixed quantum
state).

In summary: by tuning the PMD, we can move from weak to strong
measurements of polarization; the PDL performs the post-selection
of a pure or of a mixed state of polarization. Any telecom
network, devices connected by fibers, is performing "weak
measurements with post-selection". Just as in the example of
quantum cloning discussed above, all this can be (and {\em is
actually}) described by the classical theory of light.

\subsection{The fundamental role of entanglement}

We have shown that two results thought to be "typically quantum",
namely the no-cloning theorem and the theory of weak measurements,
can be demonstrated with classical light and standard telecom
devices. The key for a deep understanding is the conceptual
distinction between {\em two superposition principles}: the
classical one, which is dynamical (fields superpose because
Maxwell's equations are linear); and the quantum one, which is
kinematical: states are superposed. These two superposition
principles, at the level of interpretation, have a completely
different meaning. However, it may difficult to tell which is
acting in a real situation.

I would like to extend this observation to stress the fundamental
role of {\em entanglement}. In the traditional textbooks of
quantum mechanics, entanglement has been considered a kind of a
side-issue, and in any case a derived notion: if a composed
quantum system is described by a tensor product of Hilbert spaces,
and if the superposition principle has to hold in this total
space, then non-factorizable states must appear. In other words,
traditionally one starts with the quantum physics of the single
system, states the superposition principle in this context, and
derives the existence of entanglement a posteriori. While this may
be an unavoidable approach for a didactic course, I don't think
that the view so conveyed is really the whole story. Students
meeting the Stern-Gerlach experiment in their "quantum physics"
course fail to realize that they have studied its analog with
light polarization some months before, in their lectures on "{\em
classical} electrodynamics". But what does it mean? Is a spin
$\demi$ classical? Or is polarization a quantum intruder in the
classical theory of light?

The solution comes by noticing that the Stern-Gerlach experiment
is not the only experimental result involving the spin! The spins
of the electrons explain the Mendeleev table via the Pauli
exclusion principle, a principle that has no classical analogue;
different cross-sections have been observed in scattering
experiments, according to whether the full spin was in a symmetric
or in the anti-symmetric state; spins couple coherently to one
another in nuclear magnetic resonance, or to the polarization of
photons in atomic physics... The list may rapidly become very
large. But if we give a second glance to this list, we notice that
it contains only phenomena in which {\em two or more} quantum
systems are involved. And if we finally notice that "coherent
interaction" means "entanglement", we have the solution: {\em we
know that a single spin $\demi$ is a quantum object because we
observe the consequences of its entanglement with other spins or
other degrees of freedom}. The same can be said for polarization.

The difference between the "classical superposition principle of
waves" and the "quantum superposition principle of states" lies in
the fact that only the second gives rise to entanglement. If we'd
have only the quantum physics of the single particle (the
Stern-Gerlach experiment, Young's double-slit...), the most
economic solution would be to adopt once for all the de
Broglie-Bohm view of a real particle guided by a hidden wave ---
and we'd lose all the fascinating view of the world that is
inspired by quantum physics.

\section{Conclusion}

The main message I wanted to convey is that "classical" and
"quantum" physics --- or information --- are tightly connected.
Specifically, I have discussed how in the analysis of the security
of quantum cryptography, we discover numbers that come from the
analysis of the security of classical cryptography (Section 2);
and how experiment with classical light and standard telecom
devices can provide demonstrations of the no-cloning theorem and
of the theory of weak measurements with post-selection (Section
3).

In this text, I reported on results obtained at the University of
Geneva under the direction of prof. Nicolas Gisin, together with
Antonio Ac\'{\i}n, Nicolas Brunner, Daniel Collins, Sylvain Fasel,
Gr\'egoire Ribordy and Hugo Zbinden. I also benefited from several
discussions with Fran\c{c}ois Reuse and Antoine Suarez.

\end{document}